\documentclass[10pt,conference]{IEEEtran}
\IEEEoverridecommandlockouts
\usepackage{cite}
\usepackage{amsmath,amssymb,amsfonts}
\usepackage{algorithmic}
\usepackage{graphicx}
\usepackage{textcomp}
\usepackage{float}
\usepackage{xcolor}
\usepackage{comment}
\def\BibTeX{{\rm B\kern-.05em{\sc i\kern-.025em b}\kern-.08em
    T\kern-.1667em\lower.7ex\hbox{E}\kern-.125emX}}
\usepackage{tabularx}
\usepackage{array}
\usepackage[table]{xcolor}
\usepackage{xurl} 
\usepackage{url}

\newcolumntype{Y}{>{\raggedright\arraybackslash}X} 
\newcolumntype{C}[1]{>{\centering\arraybackslash}p{#1}}
\newcolumntype{L}[1]{>{\raggedright\arraybackslash}p{#1}}

\usepackage{pbalance}

\makeatletter
\let\OldIEEEauthorblockN\IEEEauthorblockN
\renewcommand{\IEEEauthorblockN}[1]{\OldIEEEauthorblockN{\small #1}}
\let\OldIEEEauthorblockA\IEEEauthorblockA
\renewcommand{\IEEEauthorblockA}[1]{\OldIEEEauthorblockA{\scriptsize #1}}
\newcommand{\linebreakand}{%
  \end{@IEEEauthorhalign}
  \hfill\mbox{}\par
  \mbox{}\hfill\begin{@IEEEauthorhalign}
}
\makeatother
\begin{document}
\title{Using LLMs to Elicit Security Requirements for Service-Oriented Cyber Ranges}
%SecureRange: A Semi-Automated Security Requirements Eliciation Process
% Using LLMs to Elicit Security Requirements for Service-Oriented Cyber Ranges

\author{
\IEEEauthorblockN{
Michail Takaronis,
Athanasia Kollarou,
Georgios Kavallieratos,
Vasileios Gkioulos, and
Sokratis Katsikas
}
\IEEEauthorblockA{
Dept. Information Security and Communication Technology\\
Norwegian University of Science and Technology\\
Gjøvik, Norway\\
\{michail.takaronis, athanasia.kollarou, georgios.kavallieratos,\\
vasileios.gkioulos, sokratis.katsikas\}@ntnu.no
}
}

\maketitle

\begin{abstract}
Cyber ranges are complex environments comprising many interacting components and stakeholders with different security concerns. The Service-Oriented Cyber Range (SOR) is no exception, particularly when it comes to training scenarios targeting critical infrastructure. Security concerns are translated into security requirements, the elicitation of which is usually difficult and time-consuming. This work examines how large language models can assist in eliciting security requirements for a service-oriented range and help produce a useful baseline for designers and developers.
The approach follows a SEBoK-guided process in which security mission objectives and stakeholder needs were first identified and then provided as a prompt context along with architectural guidelines to five LLMs: GPT-5.2, Gemini 3.1 Pro, Grok 4.1, Sonar, and Kimi K2.5. The models generated 84 security requirements in total, which were consolidated into a comprehensive set of 27 requirements and then mapped to the architectural layers of the service-oriented range. The final set was evaluated by five cybersecurity experts against the criteria of necessity, clarity, completeness, feasibility, and testability, with an additional rejection option. The results showed a high acceptance rate, specifically for necessity with 98.5\%, clarity with 87.4\%, completeness with 85.2\%, feasibility with 78.5\%, and rejection with 0.7\%. Testability was lower at 44.4\%, indicating a slight lack of information on how these requirements could be tested. These findings show that LLMs can support early stages of the elicitation of security requirements, although human review is still needed, especially to improve or adjust certain aspects of the requirements.
\end{abstract}


\section{Introduction}

The increasing complexity and interconnection of critical infrastructures have elevated the importance of robust cybersecurity practices, particularly in environments designed for training, testing, and experimentation, such as cyber ranges (CRs) \cite{5,13}. A Service-Oriented Range (SOR) is a cyber range designed according to service-oriented architecture (SOA) principles, where the CR capabilities are given as services across functional layers \cite{sor}. The distinction from a traditional CR lies in the infrastructure design, scenario distribution, and platform function. A SOR promotes modularity, scalability, service reuse, and interoperability across the platform layers.  In order to strengthen the cybersecurity posture of the SOR, it is necessary to define a secure architecture. Acknowledging the fact that the SOR combines nine architectural layers, services, and cross-layer dependencies, it is essential to establish a structured process for deriving security requirements that will ensure a secure system and enable tracking of which system needs are covered.

Security requirements elicitation is a foundational activity in systems engineering, translating stakeholder needs and mission objectives into actionable system specifications. However, traditional elicitation methods rely on manual activities, such as interviews, workshops, questionnaires, and document analysis, which shows that the process is as iterative and often constrained by time and cost \cite{ZowghiCoulin2005,PachecoGarciaReyes2018}. Recent advances in Large Language Models (LLMs) present new opportunities to support and enhance this process by automating parts of the requirements generation and analysis workflow.

In this context, this paper investigates the use of LLMs to assist in the elicitation of security requirements for an SOR. Building upon established systems engineering practices and on the cyber range literature, we propose a structured methodology that integrates security mission objectives, stakeholder needs, and architectural considerations into LLM-driven security requirements generation. Multiple LLMs are leveraged to elicit security requirements; these are subsequently consolidated into a unified and coherent set of requirements.

The main contribution of this work lies in proposing a methodology that leverages LLMs to support the systematic derivation of security requirements in service-oriented cyber ranges, while maintaining traceability to stakeholder needs and architectural layers. Furthermore, the resulting requirements provide a practical baseline for designers and developers of cyber ranges, contributing to more secure and resilient platform implementations.

The remainder of this paper is organized as follows. Section II reviews related work, and Section III presents the proposed methodology for deriving security requirements for an SOR. Section IV defines the identified security mission objectives, while Section V describes the security stakeholder needs. Section VI details the LLM-assisted security requirements elicitation process and the consolidation of results. Section VII maps the derived security requirements to the architectural layers of the SOR. Section VIII discusses the findings, limitations, and evaluation results, and finally, Section IX concludes the paper and outlines future work.

\section{Related Work}

Several works in the literature have explored the key functionalities and capabilities of Cyber Ranges (CRs).  ECSO, in \cite{ref9}, identified the key features of a CR platform to support best practices and develop guidelines for CR development. Features of CR, such as technical components, realism and fidelity, access considerations, scalability and elasticity, and curriculum and learning outcomes are provided in \cite{ref6}. Additionally, the key functionalities and capabilities of CRs are provided in \cite{kampourakis2025step}. The analysis therein identified, in addition, the functions and information exchanged between different components of the reference architectural model for a CR. The capabilities, roles, and tools for CRs are reviewed in \cite{yamin2020cyber}.  In \cite{lazarov2025lessons}, lessons learned from extensive testing of cybersecurity scenarios with students in different environments are discussed, and the perspectives of the teacher, university student, and secondary school student in a CR environment are explored.  The platform requirements, high-level architecture, and the user interface and interaction of the KYPO CR, as well as  th main capabilities of the CRF are presented in \cite{vykopal2017kypo}. The Cyber Range Design Framework is proposed in \cite{katsantonis2023cyber}, focusing on the architectural aspects of the cyber range. Further, the functions and goals of the components are discussed. A taxonomy of the scenarios, environment, econometrics, teaming, learning, monitoring, management, and technology aspects of CRs is proposed in \cite{lillemets2025systematic}. 

The security aspects of CRs are explored in \cite{noponen2022cybersecurity}. By analyzing the main functions of the CR and the different use cases, several security threats, e.g.,  deployment errors, physical threats, sensitive data leakage, communication threats, cloud threats, Incident Response Plan (IRP) threats, virtual machine threats, and container threats, are identified therein.

Several security requirements engineering methods exist, and several works have compared methods, tools, and frameworks for security requirements elicitation \cite{MeadSecReq,  Nhlabatsi2010SecurityRE, pattakou2017security,kavallieratos2020shipping}. A tool that facilitates the requirements engineering process is proposed in \cite{pandey2018towards}, focusing on functional and non-functional requirements. 

Several surveys have explored the application of LLMs in the requirements engineering process, highlighting the advantages, limitations, and available techniques \cite{khan2025large,zadenoori2025large,melo2025investigating}. A systematic literature review \cite{ali2025prompting} explores prompting strategies used with LLMs for requirements elicitation. In \cite{sor}, the SOR reference architecture was introduced, showing and providing guidance on how CR capabilities and functionality can be organized into functional layers following the SOA principles. However, the elicitation of security requirements for an SOR, leveraging LLMs, has not, to the best of our knowledge, been explored. 

This paper leverages prior research on security requirements engineering methodologies, tools for requirements elicitation, and studies on the application of LLMs in requirements engineering. It also draws on existing work on CR architectures and requirement identification to contextualize and guide the elicitation of security requirements for a SOR, by proposing a structured, SEBoK-guided process to identify security requirements for a SOR.

\section{Methodology}
\begin{figure*}[!ht]
\centering
\includegraphics[width=0.65\textwidth]{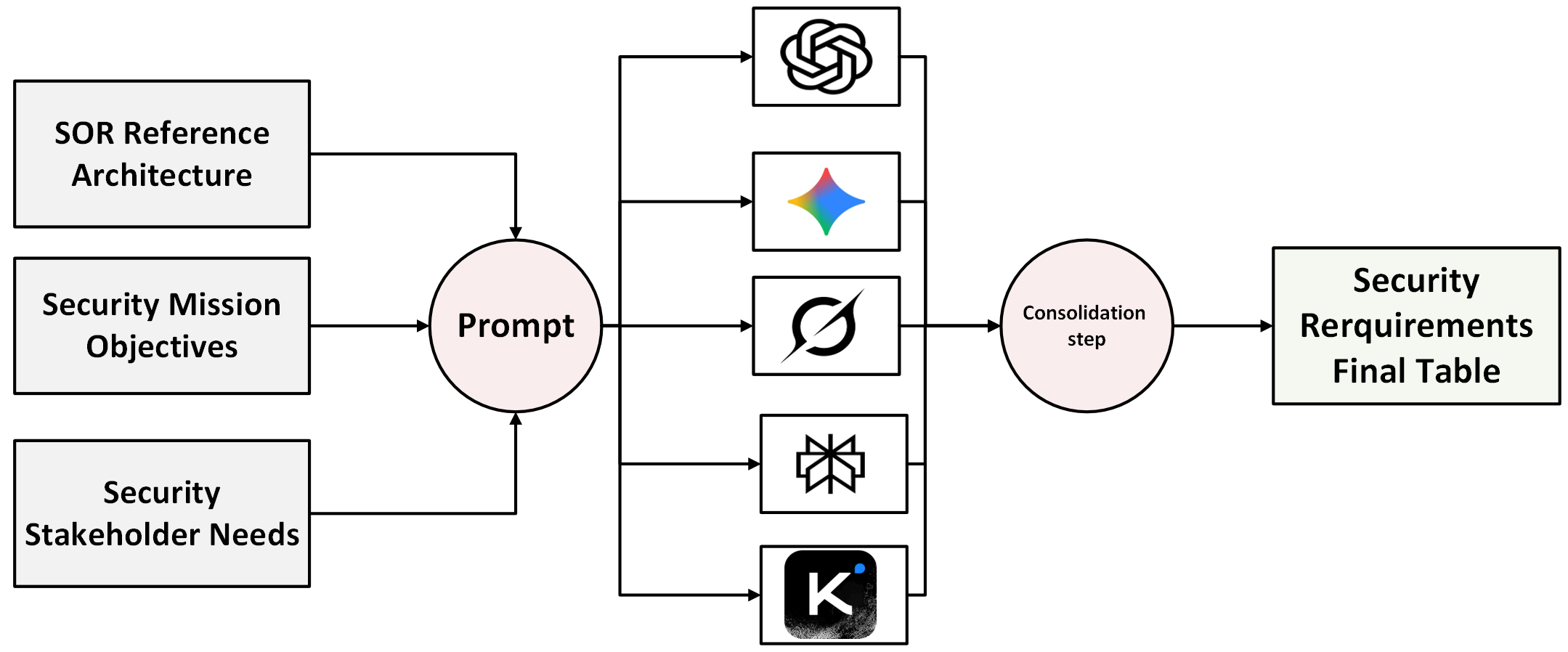}
\caption{Methodology}
\label{fig:meth}
\end{figure*}

% SEBoK + LLM for requirements
According to SEBoK \cite{sebok}, system requirements are derived by translating mission objectives and stakeholder needs into technical specifications of what the system must do to satisfy those needs, which is our final goal. The proposed process is illustrated in Fig.~\ref{fig:meth}, and details on its four phases are presented in the sequel.

\begin{enumerate}
    \item \textbf{Identify security mission objectives:} First, we analyze existing guides, taxonomies, and reference architectures for cyber ranges to identify the security mission goals (SMOs) of an SOR. From these sources, we extract a set of high-level security objectives.
    \item \textbf{Elicit security stakeholder needs:} Second, we identify the key security stakeholders and stakeholder security needs (SSNs) for a SOR. For each stakeholder class, we identify the security core needs, respecting the structural guidelines given in the SOR reference architecture \cite{sor}.
    \item \textbf{Configure the LLM task:} Thirdly, we combine the SMOs, the SSNs, and the SOR reference architecture \cite{sor} into a contextualised package in the format of an LLM prompt. This package is then fed into five LLMs using a fixed structure and role instruction that casts the model as a security requirements engineer for an SOR. The package includes the seven SMOs, the eight SSNs and a condensed description of the nine-layer SOR reference architecture. The output is an explicit contract that specifies a six-column schema (ID, name, description, verification measure, class and sources), as well as a constraint that requires the relevant SMOs/SSNs to be derived and cited for each requirement in order to preserve traceability \cite{experiment}.
    \item \textbf{Consolidation and Requirement Specification:} Finally, we perform a manual consolidation step to transform the five LLM tables into a unified set of SOR security requirements, by following a set of rules, presented in detail in Section \ref{LLM-Assisted Security requirements}. 
\end{enumerate}

Once the security requirements have been derived, we map each requirement to the most relevant SOR layer(s), and we capture the resulting layer-to-layer relationships in a connected UML dependency diagram. To verify the set of final security requirements, five cybersecurity experts were invited to review and evaluate them.

\section{Security Mission Objectives for the Service‑Oriented Cyber Range}

Mission objectives in the context of systems engineering refer to the high-level, actionable goals that must be achieved to fulfill the overall purpose or mission of a system \cite{sebok}. Security Mission Objectives (SMOs) describe the general security goals that the system should fulfil to ensure its security. These are used to define the security requirements.

By leveraging the analysis of threats identified in  \cite{noponen2022cybersecurity}; the capabilities, roles, and tools for CRs as reviewed in \cite{yamin2020cyber}; the analysis of the key cyber range audiences, capabilities, and features and the proposed common set of features and the terminology in \cite{ref9}; and the lessons learned  from extensive testing of cybersecurity scenarios as reported in \cite{lazarov2025lessons},  we identify the following security mission objectives for the SOR that adheres by the SOA principles:   

\begin{itemize}
    \item \textbf{SMO1:} The architecture has to ensure that all exercises and experiments are executed in a controlled environment that prevents unintended impact on external systems and ensures legality and safety of training and testing activities.
    \item \textbf{SMO2:} All communications, both service‑to‑service and user‑to‑service, must be protected against interception and manipulation.
    \item \textbf{SMO3:} The architecture must preserve confidentiality and integrity of all platform‑managed data, including user identities, exercise configurations, monitoring logs/telemetry, orchestration metadata, and governance policies.
    \item \textbf{SMO4:} The architecture must ensure trustworthy identity and access management for all user roles and for service identities.
    \item \textbf{SMO5:} Ensure that teams cannot unintentionally share or interfere with the same scenario execution context to prevent cross-team disclosure and contamination.
    \item \textbf{SMO6:} Ensure that the platform can detect and report security-relevant conditions affecting the platform itself with sufficient traceability to support response and assurance.
    \item \textbf{SMO7:} Ensure that, when operating as part of a federated cyber range ecosystem, the platform enforces boundary protections and can monitor/control suspicious or policy-violating behaviour originating from federation links or partner environments.
\end{itemize}

\section{Security Stakeholder needs for the Service‑Oriented Cyber Range}

Several works in the literature have examined the audiences, use cases, and feature sets of cyber ranges, providing a basis for identifying the needs of security stakeholders in an SOR. The ECSO Cyber Range Features Checklist \cite{ref9} provides a structured set of capabilities and participant roles that can be used by stakeholders to formulate requirements and selection criteria when selecting or procuring CR platforms and services. These stakeholders include academia, research centres, government agencies, and industry \cite{ref9}. The NIST Cyber Range Guide includes different types of audiences, such as educators, trainees, and organizations seeking skills development, validation, or situational testing \cite{ref6}. Furthermore, it provides an overview of the features that meet their needs, such as orchestration, learning management, assessment, and monitoring \cite{ref6}. Reference architectures and taxonomies for CRs further emphasize the roles of red, blue, white, and green teams, as well as operators and decision-makers \cite{kampourakis2025step,yamin2020cyber,katsantonis2023cyber,lillemets2025systematic}. They also highlight the aspects that shape stakeholder expectations, namely scenarios, environments, teaming, learning, monitoring, and management. By analyzing these, the following Security Stakeholder Needs (SSNs) are identified:

\begin{itemize}
    \item \textbf{SSN01 - Platform administrators / Green Team}: Platform administrators require the ability to operate and evolve the service‑oriented cyber range without exposing control planes or management tooling to unnecessary risk \cite{ref9, yamin2020cyber}. This implies a strong isolation between management services and tenant environments, hardened orchestration interfaces, and carefully implemented authorization for administrative functions in line with identified architectural components and management features \cite{kampourakis2025step,noponen2022cybersecurity}.
    \item \textbf{SSN02 - Instructors}: Instructors need secure but user-friendly access to orchestration, learning management (LMS), and monitoring services, to enable them to configure, launch, and supervise exercises \cite{kampourakis2025step}. This can be achieved with clear role setting that protects learner data, assessment results, and internal platform analytics from unnecessary disclosure \cite{katsantonis2023cyber}.
    \item \textbf{SSN03 - Exercise designers / White Team}: Exercise designers require mechanisms to define and implement least-privilege access policies within scenarios \cite{yamin2020cyber, katsantonis2023cyber}. These policies cover network reachability, credentials, and data flows. They are designed to prevent learners from accessing the underlying infrastructure or leaking sensitive content beyond the intended exercise boundaries.  \cite{vykopal2017kypo,noponen2022cybersecurity}.
    \item \textbf{SSN04 - Learners / Red and/or Blue Team}: Learners require execution environments that safely contain offensive and defensive tools, prevent collateral damage to external systems, and ensure that identity and performance data are collected, stored, and shared in accordance with the defined privacy policies \cite{ref6,lazarov2025lessons,noponen2022cybersecurity}.
    \item \textbf{SSN05 - Security researchers}: Security researchers require environments in which they can execute potentially novel or disruptive techniques under strong containment \cite{ref9,ref6}. These environments require controlled mechanisms for importing assets and exporting data to prevent the leakage of confidential datasets, proprietary tools, or discovered vulnerabilities, while enabling realistic experimentation and observability \cite{yamin2020cyber,noponen2022cybersecurity}.
    \item \textbf{SSN06 - Federation partners}: Federation partners require trustworthy interoperability services that enforce strict boundary controls, authenticate and authorise cross-range interactions, and provide the capability to monitor, constrain, or revoke federated links \cite{kampourakis2025step,katsantonis2023cyber,ref23}. This is to ensure that behaviour from partner environments does not violate agreed technical or organisational security policies. 
    \item \textbf{SSN07 - Auditors}: Auditors require end-to-end traceability of security-relevant events across services. This includes the logging of administrative actions, lifecycle operations, and federation activities in a way that makes it clear how architectural decisions satisfy objectives over time \cite{kampourakis2025step,katsantonis2023cyber,noponen2022cybersecurity}.
    \item \textbf{SSN08 - Cloud and infrastructure providers}: Cloud and infrastructure providers require details and assurances that workloads running on their services remain contained within the agreed logical and physical boundaries and comply with acceptable use \cite{ref9,yamin2020cyber,noponen2022cybersecurity}.
\end{itemize}

\section{LLM-Assisted Security requirements}
\label{LLM-Assisted Security requirements}

Following SEBoK \cite{sebok},  system requirements are derived by translating mission objectives and stakeholder needs into technical specifications for what the system must do to satisfy those needs. In this study, our goal was to elicit security system requirements for an SOR. After defining the security mission objectives, security stakeholder needs, the SEBoK methodology, and the SOR reference architecture introduced in \cite{sor}, we provided the same contextual package to five LLMs and prompted them to produce candidate security requirements. Specifically, the LLMs were tasked with deriving security system requirements. Each model was instructed to output a structured set of security requirements using the common schema, shown in \ref{tab:tab1}:
\begin{itemize}
    \item \textbf{Unique ID}: A sequential unique identifier for each requirement (e.g. SEC-01).
    \item \textbf{Requirement name}: A unique name for each requirement.
    \item \textbf{Description}: Brief textual description of the requirement.
    \item \textbf{Verification measure}: Proposed method for verifying the requirement.
    \item \textbf{Class:} A broader thematic security category.
    \item \textbf{Sources:} SMOs and SSNs that were taken into consideration to generate the security requirement.
\end{itemize}

In April 2026, all five models were queried via their public web interfaces under their default decoding settings, using identical input material. To reflect typical practitioner use, a single query was issued per model, outlining the input, output, and format of the generated result table per model. The exact model build identifiers, access dates, and the complete system and user prompts are archived in the companion repository \cite{experiment}. The models were GPT-5.2
(OpenAI, Dec 2025) \cite{openai_gpt52_2025}, Gemini 3.1 Pro (Google, Feb 2026) \cite{google_gemini31pro_2026}, Grok 4.1 (xAI,
Nov 2025) \cite{xai_grok41_2025}, Sonar (Perplexity, on Llama 3) \cite{perplexity_sonar_2026}, and Kimi K2.5 (Moonshot AI,
Jan 2026) \cite{moonshot_kimik25_2026}.

% Include the criteria of requirements from SEBoK - https://sebokwiki.org/wiki/System_Verification
Each model received the same input material and was asked to generate security requirements in the target table format. After collecting the results, we performed a consolidation step to create a unified requirement set in a table structure, Table~\ref{tab:tab1}. The reason for this was that the LLMs produced overlapping requirements, with different names and slightly different wording. Consolidation was carried out using the following rules:
\begin{itemize}
    \item \textbf{Intent-based deduplication:} Two entries were treated as duplicates if they had the same underlying security control objective. For instance, \textit{Isolation} and \textit{Tenant isolation} were considered the same, as they targeted cross-team interference prevention.
    \item \textbf{Preference for specificity:} When merging duplicates, we kept the most precise or strongest phrasing requirement.
    \item \textbf{Decomposition vs. fusion:} Some generated requirements combined more than one security concern or control objective (e.g., a single requirement addressing both confidentiality and integrity). These requirements were split and aligned with the appropriate classes. For example, Data Integrity is absorbed in Integrity.
    \item \textbf{Verification method, class, and source alignment:} Some models proposed different verification methods, classes, and sources. We aligned verification methods and classes to the most appropriate and reusable ones by integrating specific choices into broader categories, where the specifics can be implied. For instance, \textit{protocol analysis}, \textit{network testing}, and \textit{isolation testing} were aligned to broader methods such as \textit{config review}. In terms of sources, we retained the SMOs and SNNs that appeared in the results of the majority of models and were most clearly linked to the requirements.
\end{itemize}

\begin{table*}[!ht]
\caption{SOR Security Requirements}
\label{tab:tab1}
\centering
\begin{scriptsize}
\rowcolors{2}{gray!10}{white}
\begin{tabularx}{\textwidth}{|c|C{2.1cm}|C{7cm}|c|C{2cm}|C{1.65cm}|}
\hline
\textbf{ID} & \textbf{Requirement} & \textbf{Description} & \textbf{Verification} & \textbf{Class} & \textbf{Sources}\\
\hline
SEC-001 & Containment & The platform shall enforce network/system containment and egress controls to prevent exercise activity/traffic/tooling from impacting external environments and to support safe/legal operation (including provider safety assurances). & Config review & Containment/Isolation & SMO1, SSN04, SSN05, SSN08 \\
\hline
SEC-002 & TenantIsolation & The platform shall strictly isolate exercise execution contexts between exercises/teams to prevent cross-team interference, disclosure, or contamination; each exercise/team environment must be network-isolated from other exercises and from external systems. & Penetration testing & Containment/Isolation & SMO5, SSN03, SSN04 \\
\hline
SEC-003 & Encryption & The platform shall encrypt all service-to-service and user-to-service communications using mutually authenticated TLS with approved cipher suites to prevent interception and manipulation. & Penetration testing & Communications Security & SMO2, SSN01, SSN02, SSN06 \\
\hline
SEC-004 & Integrity & The platform shall protect communications and platform-managed artifacts (configs, orchestration metadata, telemetry, logs, and policies) against unauthorized modification using cryptographic protections where appropriate. & Log inspection & Data Protection & SMO2, SMO3, SSN01, SSN07 \\
\hline
SEC-005 & Confidentiality & The platform shall ensure confidentiality of platform-managed data (identities, configurations, policies, logs, learner performance, and telemetry) via encryption-at-rest and role-restricted access. & Config review & Data Protection & SMO3, SSN02, SSN04, SSN07 \\
\hline
SEC-006 & Authentication & The platform shall authenticate all users and services against a centrally managed identity provider before granting platform access; privileged roles should use multi-factor authentication. & Integration testing & IAM & SMO4, SSN01, SSN02, SSN06 \\
\hline
SEC-007 & Authorization & The platform shall enforce role-based access control aligned with stakeholder responsibilities and least-privilege across platform services and administrative functions (including simulated users and service identities). & Penetration testing & IAM & SMO4, SSN01, SSN02, SSN03 \\
\hline
SEC-008 & LeastPrivilege & The platform shall ensure all user and service identities operate with least privilege permissions. & Access review & IAM & SMO4, SSN01, SSN03 \\
\hline
SEC-009 & IdentityFederation & The platform shall support secure federation of identities with partner cyber ranges. & Integration testing & Federation & SMO7, SSN06 \\
\hline
SEC-010 & NetworkSegmentation & The platform shall segment internal networks to separate infrastructure, management, and exercise traffic; the management plane must be segmented and isolated from exercise data planes. & Config review & Containment/Isolation & SMO1, SSN01, SSN08 \\
\hline
SEC-011 & Monitoring & The platform shall continuously monitor security-relevant events and detect/surface anomalous or policy-violating behaviour affecting platform components and services, with traceability to support response (e.g., by green teams). & Integration testing & Monitoring/Audit & SMO6, SSN01, SSN07 \\
\hline
SEC-012 & Logging & The platform shall generate tamper-resistant (tamper-evident), timestamped, attributable logs of security-relevant events to support response, assurance, and auditing; logs should be immutable where feasible. & Log inspection & Monitoring/Audit & SMO6, SSN07 \\
\hline
SEC-013 & Traceability & The platform shall maintain traceability between actions, identities, and platform events for auditing and accountability. & Audit review & Monitoring/Audit & SMO6, SSN01, SSN07 \\
\hline
SEC-014 & Alerting & The platform shall surface security alerts for suspicious or policy-violating behaviour. & Integration testing & Monitoring/Audit & SMO6, SSN01, SSN07 \\
\hline
SEC-015 & BoundaryProtection & The platform shall enforce boundary controls on all federation and external connectivity interfaces; federation links must enforce mutual authentication and boundary policy inspection, and monitor/block suspicious or policy-violating behaviour from federation links. & Log inspection & Federation & SMO7, SSN06, SSN08 \\
\hline
SEC-016 & PolicyEnforcement & The platform shall enforce governance and security policies across all platform services and evaluate/enforce inter-range governance policies on federation partner interactions in real time. & Policy review & Governance & SMO3, SMO4, SMO7, SSN01, SSN06 \\
\hline
SEC-017 & DataExportControl & The platform shall restrict and audit export of platform-managed/exercise data to external destinations; exports require explicit role-authorized approval and produce immutable audit records. & Audit review & Data Protection & SMO3, SMO5, SSN05, SSN07 \\
\hline
SEC-018 & AdministrativeSecurity & The platform shall secure privileged administrative interfaces against unauthorized access or misuse; privileged actions must be non-repudiably attributed to a verified identity and persisted in immutable audit records. & Log inspection & Monitoring/Audit & SMO4, SMO6, SSN01, SSN07 \\
\hline
SEC-019 & Compliance & The platform shall enforce controls ensuring exercise operations comply with legal and provider requirements, and provide assurance evidence to providers that traffic/tooling use remains contained and compliant. & Auditing & Compliance & SMO1, SSN07, SSN08 \\
\hline
SEC-020 & ServiceIdentity & Each platform microservice must hold a cryptographically verified identity used for mutual authentication when invoking other platform services. & Config review & IAM & SMO2, SMO4, SSN01 \\
\hline
SEC-021 & SecretManagement & All cryptographic keys and service credentials must be stored, distributed, and rotated using a dedicated secrets management service. & Config review & IAM & SMO3, SMO4, SSN01 \\
\hline
SEC-022 & SessionManagement & User sessions must enforce configurable idle timeouts and support immediate revocation by platform administrators. & Integration testing & IAM & SMO4, SSN01, SSN02 \\
\hline
SEC-023 & Patching & Platform software components must be continuously vulnerability-scanned and follow a defined, auditable patch management lifecycle. & Config review & Vulnerability Management & SMO6, SSN01 \\
\hline
SEC-024 & Fidelity & The platform must support high-fidelity experimentation/training while maintaining strict containment and safety controls. & Penetration testing & Containment/Isolation & SMO1, SMO5, SSN04, SSN05 \\
\hline
SEC-025 & Usability & The platform must provide secure, role-appropriate usability for instructors/administrators without exposing sensitive learner data or platform internals. & Usability testing & Usability & SMO4, SSN01, SSN02 \\
\hline
SEC-026 & RedTeamSafety & The platform must ensure red-team and blue-team environments protect participants from collateral damage during exercises. & Penetration testing & Containment/Isolation & SMO1, SMO5, SSN03, SSN04 \\
\hline
SEC-027 & ExerciseControl & The platform must allow the White Team to define/enforce least-privilege simulated user capabilities and prevent pivoting into range infrastructure. & Penetration testing & Exercise Governance & SMO4, SMO5, SSN01, SSN03 \\
\hline
\end{tabularx}
\end{scriptsize}
\end{table*}

\section{Mapping Security Requirements to the Service-Oriented Cyber Range Architecture}
After identifying the security requirements for the SOR, to make these requirements addressable at the architectural level, we performed some additional actions.

\begin{itemize}
    \item We formalized the SOR as a UML dependency model among SOA operational layers \cite{sor}.
    \item We mapped each security requirement to the layer(s) that provide the most appropriate implementation or enforcement point for it, based on the characteristics of the SOR \cite{experiment} and general rules for SOA architectures (OASIS SOA Reference Model) \cite{soa-arch}.
\end{itemize}

\begin{figure}[!ht]
\centering
\includegraphics[width=0.45\textwidth]{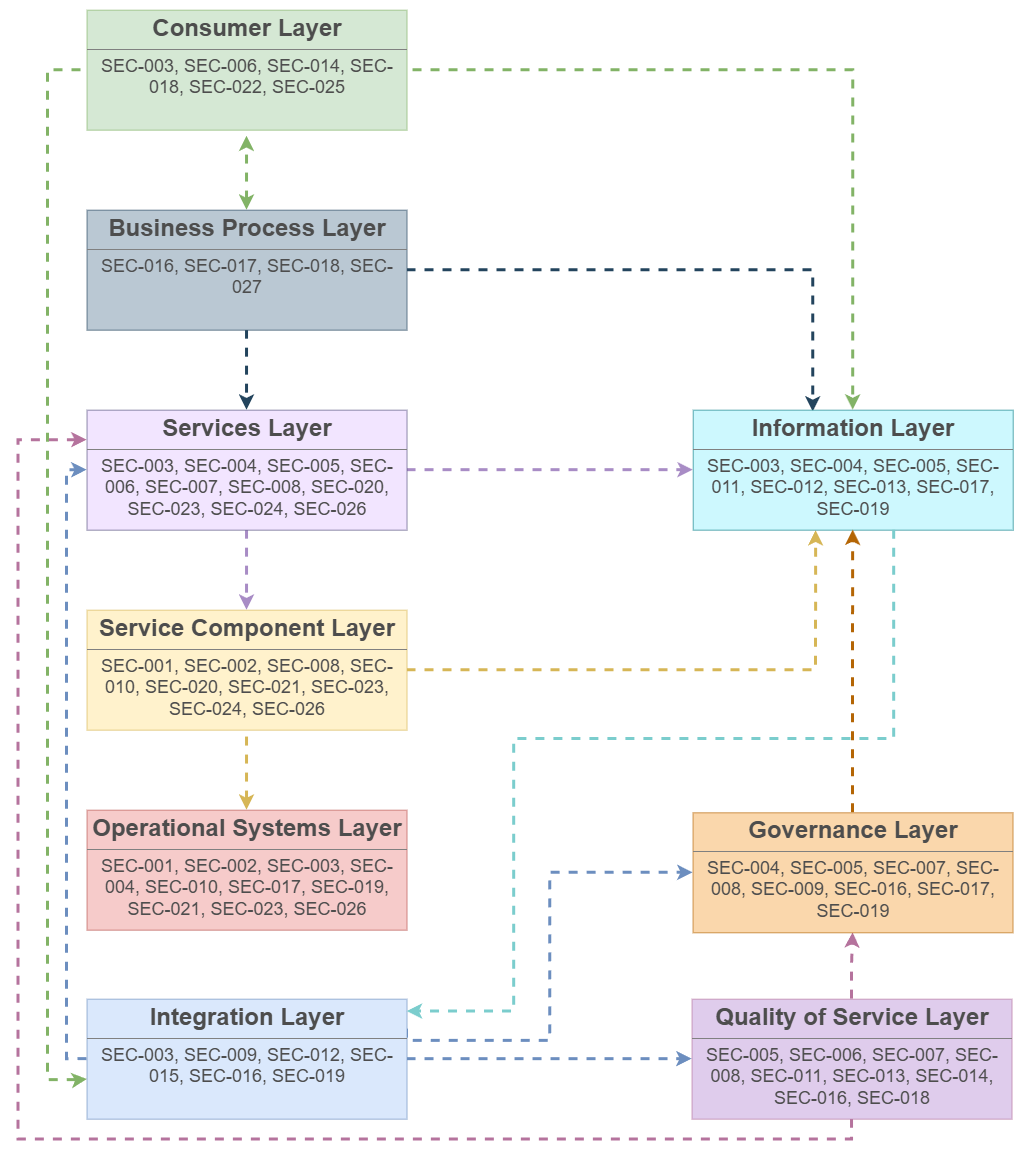}
\caption{Dependency Architecture of SOR With Mapped Security Requirements}
\label{fig:dSOCRsr}
\end{figure}

The result is presented in Fig.~\ref{fig:dSOCRsr}. The dependency arrows in the UML representation do not indicate runtime message flows. Instead, they showcase architectural dependencies among the layers. These dependencies pertain to the capabilities, metadata, or constraints required by one layer to be provided by another layer in order to enable the former to carry out its functions.

In the SOR, the \textbf{Operational Systems (OSL), Service Component (SCL), and Services layers (SL)} are responsible for core functionalities and the assurance of the smooth operation of the system. It is natural to be dependent on each other, because services are realized by deployable artifacts (VMs, Docker containers), which depend on the underlying infrastructure (Servers, Databases). Similarly, the \textbf{ Business Process layer (BPL)} depends on the Service layer because it coordinates and automates the deployment of services. For example, it manages the deployment of a virtualised environment for a cyber exercise. Moreover, this layer depends on the \textbf{Consumer (CL)} and \textbf{Information} \textbf{(IFL)} layers. From the former, it takes the actual command from the appropriate endpoint, while from the latter it fetches the appropriate data or metadata required to manage the rest of the procedures. The \textbf{Integration layer (IL)} is a type of middleware responsible for communication through message routing and transformation. User endpoints and services depend on this layer because they use it to distribute messages and data. The \textbf{Information}, \textbf{QoS}, and \textbf{Governance} layers are also dependent on this layer because of the information that will be transferred, and because the authentication/authorization of the messages relies on QoS. Finally, the policy metadata and rules that may be used in the integration layer rely on governance, creating a dependency. The SOR distinguishes that policies, business rules, and non-functional requirements are captured in \textbf{Governance (GL)}, but are monitored and enforced in the\textbf{ Quality of Service (QoS) layer}, which acts as an observer of other layers and can raise events on policy/NFR non-compliance. These operations create dependencies between the information, integration, and services layers, where data is stored, transferred, and used. Finally, the \textbf{Information layer (IFL)} aligns with the data collection and analysis modules of the ecosystem, which depend on many layers because they use these data and metadata to perform operations such as compliance, security, user identification, and exercise identification. Its dependencies on the other layers were described earlier in this section.  

After analyzing the dependencies among the SOR layers, we mapped each security requirement to the appropriate layer based on its description and the function of the layer. Security requirements focusing on isolation, network segmentation, and patching were linked to OSL and the SCL because they realize runtime, structural artifacts, and exercise isolation. Other requirements, such as boundary protection, federation interface controls, invocation-time authentication/authorization hooks, and transport-visible logging/traceability, are linked more to the IL, facilitating secure internal and external communication. QoS security requirements assignment focused on monitoring, alerting, and user authentication/authorization, while IFL is the primary evidence point for confidentiality/integrity of platform-managed data, tamper-evident logging, and traceability, because it provides the unified data. CL and BPL receive requirements related to secure user/admin interaction (sessions, privileged interfaces, usability) and exercise control/orchestration constraints, because these layers are responsible for the user interface and deployment provision within SOR. GL is the main control for defining policies and compliance objectives. Security requirements, such as integrity and confidentiality, support these policies. Finally, SL focuses more on service identities, encryption requirements, user security, and communication security, as it contains the service that comprises the SOR.

\section{Discussion}

The results show that LLM-assisted requirements elicitation can produce a diverse set of requirements, even though the sets produced by different LLMs contain many of the same requirements. Across the five models, a total of 84 security requirements were generated. GPT-5.2 and Gemini 3.1 Pro each produced 20, Sonar produced 20, Kimi K2.5 produced 14, and Grok 4.1 produced 10. All these, following the consolidation step, reduce to 27 unique requirements. This reduction from 84 to 27 entries highlights a strong convergence among models, especially on core controls such as containment, encryption, and integrity. Convergence on core controls was near-total: containment, exercise/tenant isolation, encryption, integrity, confidentiality, authentication, authorization, monitoring, and federation-boundary protection were each proposed by all five models. Differentiation came from the less-redundant models: nine of the 27 final
requirements trace to a single model, such as Alerting (GPT-5.2); ServiceIdentity, SecretManagement, SessionManagement, and Patching (Gemini); and Fidelity, Usability, RedTeamSafety, and ExerciseControl (Sonar) whereas Grok produced only the universal core and contributed no unique requirement. Consolidation merged synonymous entries (e.g., Gemini's Isolation/Segregation/Segmentation and GPT's TenantIsolation/NetworkSegmentation; the Privacy entries from Kimi and Sonar into SEC-005 Confidentiality; GPT's DataIntegrity into SEC-004 Integrity) and discarded items not traceable to a mission objective (e.g., Kimi's Availability, for which no corresponding SMO was defined) The final set of requirements was found to be logically well-suited to address the security needs across the SOR layers, providing developers and architects with a traceable baseline of the security concerns that should be addressed at each layer of the SOR.

The quality of the LLM‑generated requirements was evaluated by five cybersecurity experts, whose characteristics are briefly described in Table ~\ref{tab:experts}. They rated each of the 27 requirements against five criteria of "good" requirements, taken from the SEBoK and with definitions commonly used in other works \cite{sebok,sakhrawiSoftware2021}. The experts checked whether each requirement was \textbf{necessary, clearly described, complete, feasible, and testable}. They could also indicate that they would reject a requirement in its current form. An example entry of the evaluation form featuring the criteria checkboxes is given in \cite{experiment} under \textit{evaluation-example-entry.md}.

\begin{table}[H]
\centering
\caption{Overview of participating experts}
\begin{tabular}{ p{1.5cm} p{3.2cm} p{1.5cm}}
\hline
\textbf{Expert ID}  & \textbf{Expertise} & \textbf{Years of experience} \\
\hline
Expert1  & Cybersecurity of critical infrastructure, cybersecurity requirements engineering & 10 \\
Expert2  & Cybersecurity, AI cybersecurity, cryptography  & 20 \\
Expert3  & Cybersecurity, cyber range &   20 \\
Expert4  & Network and cloud Security, Distributed ledger technologies, AI for cybersecurity &  10  \\
Expert5  & Network security, cyber range & 5  \\
\hline
\end{tabular}
\label{tab:experts}
\end{table}

Since five experts rated 27 requirements, each criterion received 135 individual ratings. For example, the ‘necessary’ criterion appears 27 times, once for every requirement, for the 5 experts, so in total it appears 135 times. Across these ratings, the experts considered the requirements to be \textit{necessary} in 133 cases (98.5\%), indicating that they were mostly seen as relevant to the security goals. The wording was found to be \textit{clear} in 118 of the 135 ratings (87.4\%), and the requirements were considered sufficiently \textit{complete} in 115 of the 135 ratings (85.2\%). Both these scores show that the requirements were well formulated for the security goals, without any major points missing. The requirements were marked as \textit{feasible} in 106 cases (78.5\%), suggesting that they were considered realistic to implement in a lower but still high percentage. This could reveal implementation challenges, such as computational overhead or integration complexity.  However, only 60 of the 135 ratings (44.4\%) indicated it was clear how the requirements could be \textit{tested}. This indicates the need for further clarification to support the verification process. Finally, the experts indicated \textit{reject} only in one of the 135 ratings (0.7\%). The high percentages noted across most criteria indicate that the defined security requirements can serve as a baseline for stakeholders to use as a starting point when defining their own requirements. Beyond the quality of each requirement, the final set appears reasonably complete as a baseline because it covers the identified SMOs and SSNs.  

Despite the promising results, some limitations should be acknowledged. First, the requirement elicitation process depends heavily on the quality and completeness of the prompt context, including mission objectives, stakeholder needs, and reference architecture descriptions. Incomplete or biased input material could lead to incomplete requirement sets. Second, the consolidation step was performed manually, introducing a degree of subjectivity in requirement selection and alignment. We did not use an LLM for carrying out this step because this would have added another generative layer and could possibly reduce transparency in how duplicates were merged and normalized. Moreover, the manual application of the fixed rules allowed us to preserve traceability to the original outputs. Finally, the work focused on requirement identification and architectural mapping rather than empirical validation of the requirements through implementation or operational testing in a real cyber range environment.

\section{conclusion}
This work showed that LLMs can support the elicitation of security requirements for an SOR when they are guided by clearly defined security mission objectives, stakeholder needs, and architectural context. Using five LLMs, the study generated 84 candidate requirements that were consolidated into 27 unique requirements. The final set was then mapped to a UML class diagram representation of the SOR reference architecture, showcasing the dependencies between the layers and their security needs. The evaluation of the proposed set was assigned to five experts with relevant fields of expertise, and indicates that the resulting requirements are useful and relevant to the SOR security needs. The necessity, clarity, completeness, and feasibility were rated with high percentages, while testability remained the weakest criterion and the rejection rate was negligible. Thus, the study confirmed that LLMs can help define security requirements, but their output would still benefit from expert review and refinement. Limitations lie in the dependence on the quality of the given prompt, the subjectivity introduced by the manual consolidation step, and the lack of validation through implementation or operational testing in a real CR environment. These limitations mean that the presented requirements can serve as a good starting point for the requirements elicitation procedure. Our future plans focus on comparing with established security-requirements-engineering methods, automating the consolidation step to make it more efficient and accurate, as well as empirically validating the set for deployment instances of service-oriented ranges.

\section*{Acknowledgments}
This work has received funding from the Research Council of Norway through the SFI Norwegian Centre for Cybersecurity in Critical Sectors (NORCICS) project no. 310105, and  has been co-financed by the European Regional Development Fund of the European Union through the CYBERUNITY project, EU grant 101128024.

\bibliographystyle{plain}
\bibliography{references}

\end{document}